\documentclass[preprint,showpacs,titlepage,aps,prd,
amsmath,byrevtex,nofootinbib]{revtex4}

\usepackage{graphicx,amssymb}

\def\lsim{\raise0.3ex\hbox{$\;<$\kern-0.75em\raise-1.1ex
\hbox{$\sim\;$}}}
\def\gsim{\raise0.3ex\hbox{$\;>$\kern-0.75em\raise-1.1ex
\hbox{$\sim\;$}}}

\DeclareMathAlphabet{\mathsc}{OT1}{cmr}{m}{sc}

\begin{document}

\preprint{ Fermilab-Pub-05-041-T}
\preprint{hep-ph/0503283}

%%%%%%%%%%%%%%%%%%%%%%%%%%%%%%%%%%%%%%%%%%%%%%%%%%%%%%%%%%%%%%%%%%%%%%%%%%%%%%
\title{\Large Another possible way to determine\\ the Neutrino Mass Hierarchy}
%%%%%%%%%%%%%%%%%%%%%%%%%%%%%%%%%%%%%%%%%%%%%%%%%%%%%%%%%%%%%%%%%%%%%%%%%%%%%%%

\author{Hiroshi Nunokawa$^1$}
\email{nunokawa@fis.puc-rio.br}  
\author{Stephen Parke$^2$}
\email{parke@fnal.gov}
\author{Renata Zukanovich Funchal$^3$}
\email{zukanov@if.usp.br}
\affiliation{
$^1$\sl Departamento de F\'{\i}sica, Pontif{\'\i}cia Universidade Cat{\'o}lica do Rio de Janeiro, \\
C. P. 38071, 22452-970, Rio de Janeiro, Brazil \\
$^2$Theoretical Physics Department, Fermi National Accelerator Laboratory,\\  
P.O. Box 500, Batavia, IL 60510, USA\\
$^3$Instituto de F\'{\i}sica, Universidade de S\~ao Paulo, 
 C.\ P.\ 66.318, 05315-970 S\~ao Paulo, Brazil}

\date{March 29, 2005 
%******** \input /tmp/time.txt
}

\vglue 1.4cm
%%%%%%%%%%%%%%%%%%%%%%%%%%%%%%%%%%%%%%%%%%%%%%%%%%%%%%%%%%%%%%%%%
%    Abstract
%%%%%%%%%%%%%%%%%%%%%%%%%%%%%%%%%%%%%%%%%%%%%%%%%%%%%%%%%%%%%%%%%
%\hfuzz=25pt
\begin{abstract}
  We show that by combining high precision measurements of the
  atmospheric $\delta m^2$ in both the electron and muon neutrino (or
  anti-neutrino) disappearance channels one can determine the neutrino
  mass hierarchy.  The required precision is a very challenging fraction of one per cent
  for both measurements.
  At even higher precision, sensitivity to the
  cosine of the CP violating phase is also possible.  This method for
  determining the mass hierarchy of the neutrino sector does not
  depend on matter effects.
 \end{abstract}

%\vskip2pc]

\maketitle

Neutrino flavor transitions have been observed in atmospheric,
solar, reactor and accelerator neutrino experiments. 
Transitions for at least two different E/L's 
(neutrino energy divided by baseline) are seen.
To explain these transitions, extensions to the Standard Model 
of particle physics are required. 
The simplest and most widely accepted extension is to 
allow the neutrinos to have masses and mixings,
similar to the quark sector, then these flavor transitions can
be explained by neutrino oscillations.  

This picture of neutrino masses and mixings has recently come into
sharper focus with the latest salt data presented by the SNO
collaboration\cite{SNO}.  When combined with the latest KamLAND
experiment\cite{KamLAND} and other solar neutrino
experiments\cite{SK_solar,solar_other} the range of allowed values for
the solar mass squared difference, $\delta m^2_{21}$, and the mixing
angle, $\theta_{12}$, are \footnote{We use the notation of ref.\cite{sinsq} throughout.}
\begin{eqnarray}
+7.3 \times 10^{-5} {\rm eV^2} < 
& \delta m^2_{21} & 
< +9.0 \times 10^{-5} {\rm eV^2} \nonumber \\
0.25 < & \sin^2 \theta_{12}& < 0.37 
\end{eqnarray}
at the 90 \% confidence level.  
Maximal mixing, $\sin^2 \theta_{12} =0.5 $, has been ruled 
out at greater than 5 $\sigma$.
The solar neutrino data is consistent with $\nu_e \rightarrow \nu_\mu ~{\rm and/or}
~\nu_\tau$.

The atmospheric neutrino data from SuperKamiokande has changed only slight in 
the last few years\cite{SK_atm} and the latest results from the K2K 
long baseline experiment\cite{K2K} are consistent with SK.
The range of allowed values for the atmospheric mass squared difference,
$\delta m^2_{32}$ and the mixing angle, $\theta_{23}$, are
\begin{eqnarray}
1.5 \times 10^{-3} {\rm eV^2} < 
& \vert \delta m^2_{32} \vert & 
< 3.4 \times 10^{-3} {\rm eV^2} \nonumber \\
0.36 < & \sin^2 \theta_{23} & \leq 0.64  
\end{eqnarray}
at the 90 \% confidence level.  The atmospheric data is consistent
with $\nu_\mu \rightarrow \nu_\tau$ oscillations and the sign of $\delta m^2_{32}$
is unknown.  This sign is positive (negative) if the doublet of
neutrino mass eigenstates, 1 and 2, which are responsible for the
solar neutrino oscillations have a smaller (larger) mass than the 3rd
mass eigenstate. This is the mass hierarchy question.

The best constraint on the involvement of the $\nu_e$ at the atmospheric
$\delta m^2$ comes from the Chooz reactor experiment \cite{Chooz}
and this puts a limit on the mixing angle associated with 
these oscillations, $\theta_{13}$, reported as
\begin{eqnarray}
0 \leq & \sin^2 \theta_{13} & < 0.04 
\end{eqnarray}
at the 90 \% confidence level at 
$\delta m^2_{31} = 2.5 \times 10^{-3} {\rm eV^2}$. 
This constraint depends on the precise value of
$\delta m^2_{31}$ with a stronger (weaker) constraint at higher
(lower) allowed values of $\delta m^2_{31}$.

So far the inclusion of genuine three flavor effects has not been
important because these effects are controlled by the two small
parameters
\begin{eqnarray}
{\displaystyle\delta m^2_{21} \over \displaystyle \delta m^2_{32}}\approx 0.03
& \quad {\rm and/or} \quad  \sin^2 \theta_{13} \leq 0.04.
\end{eqnarray}
However as the accuracy of the neutrino data improves it will become
inevitable to take into account genuine three flavor effects including CP and T violation.
 
One of the goals of the next generation neutrino experiments is to
establish the atmospheric mass hierarchy. Many authors have studied
how to exploit matter effects in future conventional long baseline
experiments ~\cite{lbl}, in supernova explosions ~\cite{sn} or in
experiments using non conventional neutrino beams produced in a muon
collider facility~\cite{muon} to unravel the mass hierarchy.
Here we discuss how to make this determination using precision 
disappearance experiments.

Genuine three generation effects make the effective atmospheric
neutrino $\delta m^2$ measured by disappearance experiences, in
principle, flavor dependent even in vacuum and thus sensitive to the
mass hierarchy and even to the CP phase.  This observation suggests an
alternative way to access the mass hierarchy by comparing precisely
measured values for the atmospheric $\delta m^2$ in $\bar{\nu}_e \to
\bar{\nu}_e$ (reactor) and $\nu_\mu \to \nu_\mu$ (accelerator) modes.
To illuminate this rather interesting but experimentally challenging
possibility is the purpose of this paper.  
A variant of this idea, using the solar $\delta m^2$ scale, can be found in ref.\cite{Petcov}.

Assuming three active neutrinos only, the survival probability for the
$\alpha$-flavor neutrino, in vacuum, is given by
\begin{eqnarray}
  P(\nu_\alpha \to \nu_\alpha) = P(\bar{\nu}_\alpha \to \bar{\nu}_\alpha) = 1&-&  4|U_{\alpha3}|^2|U_{\alpha1}|^2 \sin^2 \Delta_{31} \nonumber \\
  &  -& 4|U_{\alpha3}|^2|U_{\alpha2}|^2 \sin^2 \Delta_{32} 
\label{alphasurvP} \\
&  -&  4|U_{\alpha2}|^2|U_{\alpha1}|^2 \sin^2 \Delta_{21}, \nonumber
\end{eqnarray}
where $\Delta_{ij} = \delta m^2_{ij} L /4E$, $\delta m^2_{ij}=m^2_i -
m^2_j$ and $U_{\alpha i}$ are elements of the MNS mixing matrix, \cite{MNS}. The
three $\Delta_{ij}$ are not independent since the $\delta m^2_{ij}$'s
satisfy the constraint, $\delta m^2_{31}=\delta m^2_{32}+\delta
m^2_{21}$.

If we define an effective atmospheric mass squared difference, 
$\delta m^2_\eta$, which depends linearly on the parameter $\eta$, as follows
\begin{eqnarray}
  \delta m^2_{\eta} \equiv \delta m^2_{31}-\eta ~\delta m^2_{21} & = &
  \delta m^2_{32}+(1-\eta)~\delta m^2_{21}  \nonumber \\
  {\rm so~that} \quad  \Delta_{\eta}=\Delta_{31}-\eta \Delta_{21} &=  &  \Delta_{32}+(1-\eta) \Delta_{21}
  =\frac{\delta m^2_{\eta} L}{4E},
\end{eqnarray}
then we can rewrite Eqn.[\ref{alphasurvP}] using the independent variables,
$\Delta_\eta$ and $\Delta_{21}$, as
\begin{eqnarray}
  1-P(\nu_\alpha \to \nu_\alpha) & =& ~~~4|U_{\alpha3}|^2(1-|U_{\alpha3}|^2)  \left[
    \sin^2 \Delta_{\eta}  \right. \nonumber \\ 
  &  & \left. \quad \quad + \{r_1 \sin^2(\eta \Delta_{21}) + r_2 \sin^2((1-\eta)\Delta_{21})\}\cos 2 \Delta_{\eta}
  \right.\nonumber  \\ & & \left.
    \quad \quad +\frac{1}{2}\{r_1\sin(2\eta\Delta_{21})-r_2\sin(2(1-\eta)\Delta_{21})\}
    \sin 2 \Delta_{\eta} \right] \nonumber \\
  &  &+~4|U_{\alpha2}|^2|U_{\alpha1}|^2 \sin^2 \Delta_{21},
\end{eqnarray}
where
\begin{eqnarray}
 \quad r_1=\frac{|U_{\alpha1}|^2}{|U_{\alpha1}|^2+|U_{\alpha2}|^2}
& {\rm and} &r_2=\frac{|U_{\alpha2}|^2}{|U_{\alpha1}|^2+|U_{\alpha2}|^2}=1-r_1.
\end{eqnarray}
Notice that the coefficient in front of $\sin 2 \Delta_{\eta}$ is
the derivative of the coefficient in front of $\cos 2 \Delta_{\eta}$, 
with respect to $\eta \Delta_{21}$, up to a constant
factor. Therefore by choosing $\eta$ so as to set the coefficient in
front of $\sin 2 \Delta_{\eta}$ to zero one also minimizes the coefficient
in front of $\cos 2 \Delta_{\eta}$.  That is, if $\eta$ satisfies
 \begin{eqnarray}
   % \tan 2\eta \Delta_{21} & = & \frac{r_2 \sin 2\Delta_{21}}{r_1+r_2\cos2\Delta_{21}}\\
   \eta & = & \frac{1}{2\Delta_{21}} \arctan \left\{
     \frac{r_2 \sin 2\Delta_{21}}{r_1+r_2\cos2\Delta_{21}} \right\} \approx r_2,
\end{eqnarray}
one minimizes the effects of both $\sin 2 \Delta_{\eta}$ and $\cos 2
\Delta_{\eta}$ terms and this $\delta m^2_{\eta}$ with $\eta \approx r_2$ {\it is} truly the
effective atmospheric $\delta m^2$, $\delta m^2_{\rm
  eff}\vert_\alpha$, measured in $\nu_\alpha$ disappearance
experiments.  The approximation $\eta=r_2$ is excellent
provided that $\Delta_{21} \ll 1$.

Using this approximate solution for $\eta$, the effective atmospheric
$\delta m^2$ for the $\alpha$-flavor is\footnote{ An alternative way
  to derive this is to notice that the first extremum, of the terms
  in Eqn.[\ref{alphasurvP}] proportional to $|U_{\alpha 3}|^2$, occurs
  when
\begin{eqnarray}
{ |U_{\alpha1}|^2 \Delta_{31} + |U_{\alpha2}|^2 \Delta_{32}
\over
|U_{\alpha1}|^2+|U_{\alpha2}|^2  } 
& = &
\frac{\pi}{2},
\end{eqnarray}
to first non-trivial order in $\Delta_{21}$.
}
\begin{eqnarray}
\delta m^2_{\rm eff}\vert_\alpha  & \equiv & \frac{ |U_{\alpha1}|^2 ~\delta m^2_{31} + |U_{\alpha2}|^2 ~\delta m^2_{32}}
{|U_{\alpha 1}|^2+|U_{\alpha 2}|^2} =r_1 ~\delta m^2_{31} + r_2 ~\delta m^2_{32},
\label{dmsqeff} 
\end{eqnarray}
then the full neutrino survival probability in vacuum, Eqn[\ref{alphasurvP}],
can be rewritten as
\begin{eqnarray}
1-P(\nu_\alpha \to \nu_\alpha) & =& ~~~4|U_{\alpha3}|^2(1-|U_{\alpha3}|^2)  \left[
\sin^2 \Delta_{\rm eff}  \right. \nonumber \\ 
&  & \left. \quad \quad + \{r_1 \sin^2(r_2 \Delta_{21}) + r_2 \sin^2(r_1\Delta_{21})\}\cos 2 \Delta_{\rm eff}
\right.\nonumber  \\ & & \left.
\quad \quad -\frac{1}{2}\{r_2\sin(2r_1\Delta_{21})-r_1\sin(2r_2\Delta_{21})\}
\sin 2 \Delta_{\rm eff} \right] \nonumber \\
&  &+~4|U_{\alpha2}|^2|U_{\alpha1}|^2 \sin^2 \Delta_{21}.
\end{eqnarray}
If the coefficients in front of the $\cos 2 \Delta_{\rm eff}$ and $\sin 2 \Delta_{\rm eff}$ terms are expanded in powers of $\Delta_{21}$,  one finds
\begin{eqnarray}
\{r_1 \sin^2(r_2 \Delta_{21}) + r_2 \sin^2(r_1\Delta_{21})\}
& = & r_1 r_2 \Delta^2_{21} + {\cal O}(\Delta^4_{21})  \nonumber \\
\frac{1}{2}\{r_2\sin(2r_1\Delta_{21})-r_1\sin(2r_2\Delta_{21})\} &=&
\frac{2}{3}r_1 r_2(r_2-r_1)\Delta^3_{21}+{\cal O}(\Delta^5_{21}),
\end{eqnarray}
and one can see clearly that all terms linear in $\Delta_{21}$ have
been absorbed into the $\Delta_{\rm eff}$ terms.  This confirms that
$\delta m^2_{\rm eff}$, Eqn[\ref{dmsqeff}], is the effective
atmospheric $\delta m^2$ to first non-trivial order in $\delta
m^2_{21}$.  Note also that the first term odd in $\Delta_{\rm eff}$
occurs with a coefficient proportional to $\Delta_{21}^3$ which, at
the first extremum, is a suppression factor of order $10^{-4}$.

To understand the physical meaning of the effective atmospheric
$\delta m^2$ it is useful to write it as follows
\begin{eqnarray}
  \delta m^2_{\rm eff}\vert_\alpha   & = & m^2_3 -  \langle m^2_\alpha \rangle_{12}, \nonumber \\[0.5cm]
  {\rm where} \quad \quad \langle m^2_\alpha \rangle_{12} & \equiv & \frac{|U_{\alpha2}|^2 m^2_2 +|U_{\alpha1}|^2 m^2_1}
  {|U_{\alpha1}|^2+|U_{\alpha2}|^2}.
\end{eqnarray}
Now $\langle m^2_\alpha \rangle_{12}$ has a clear interpretation, it
is the $\alpha$-flavor weighted average mass square of neutrino states
1 and 2.  Thus the effective atmospheric $\delta m^2$ is the
difference in the mass squared of the state 3 and this flavor average
mass square of states 1 and 2 and is clearly flavor dependent. 

\begin{figure}
\centering\leavevmode
\vglue -1.0cm
%\hglue 2.0cm
\includegraphics[width=12.cm]{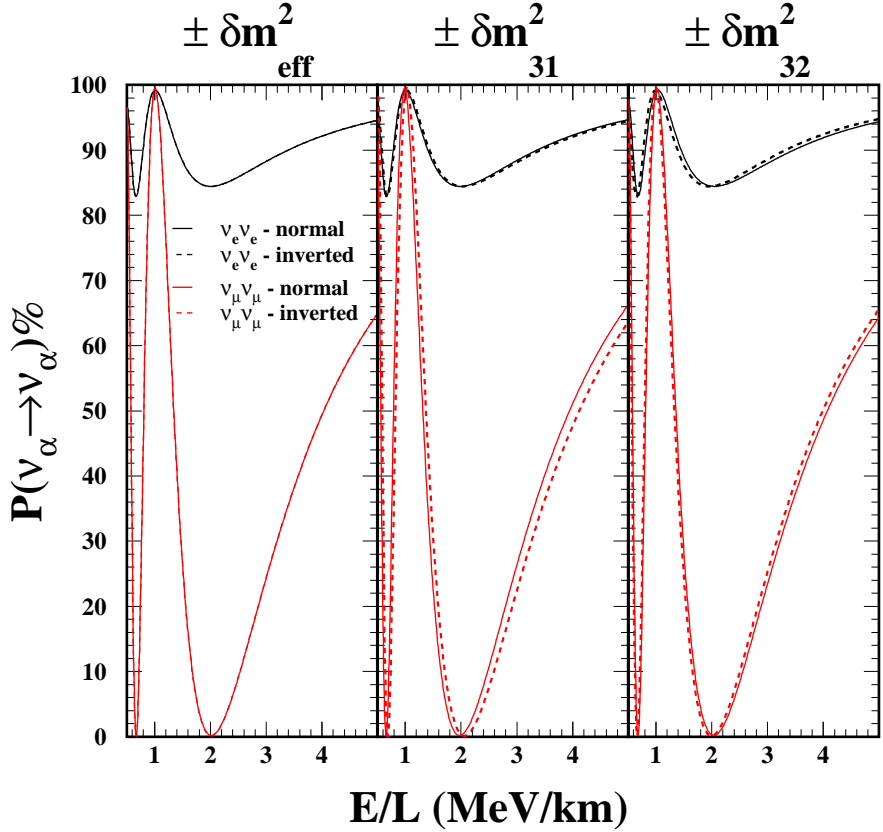}
\vglue -1.7cm
\caption{The vacuum survival probability, $P(\nu_\alpha \to \nu_\alpha)$, 
as a function of $E/L$ for the two mass hierarchies using three
  different choices of the atmospheric $\delta m^2$ whose flips sign, with constant magnitude,
  changes the hierarchy: $\delta
  m^2_{\rm eff}\vert_\alpha$ (left panel), $\delta m^2_{31}$ (middle panel) and
  $\delta m^2_{32}$ (right panel). The survival probability for the two different hierarchies 
  coincide to high precision when the effective $\delta m^2$'s, Eqn[\ref{meffe}, \ref{meffm}], are used (left panel)
  whereas they differ appreciably with the other two definitions. 
  For this figure we have used $\sin^2 \theta_{23}=0.5$ (maximal mixing),
  $\sin^2 \theta_{13}=0.04$ (Chooz bound),  $\sin^2 \theta_{12} = 0.31$,
  $\delta m^2_{21}= +8.0 \times 10^{-5} ~{\rm eV}^2$ 
   and the atmospheric $\delta m^2$ to be $2.5 \times 10^{-3} ~{\rm eV}^2$. 
 }
\label{prob}
\end{figure}

The three flavor average mass squares are\footnote{Dropping terms of order $\sin^2 \theta_{13}
\delta m^2_{21}$.}
\begin{eqnarray}
\langle m^2_e \rangle_{12} 
%&=& \sin^2\theta_{12} m^2_2 + \cos^2 \theta_{12} m^2_1\\ 
&=&\frac{1}{2}[m^2_2+m^2_1
- \cos 2 \theta_{12} ~\delta m^2_{21}] \nonumber \\
\langle m^2_\mu \rangle_{12} &=& 
%\cos^2\theta_{12}m^2_2 +\sin^2 \theta_{12}m^2_1-\delta m^2_{21} \cos \delta \sin \theta_{13} \sin 2 \theta_{12} \tan \theta_{23}  \\ & = &
 \frac{1}{2}[m^2_2+m^2_1+(\cos 2 \theta_{12}-2 \cos \delta \sin \theta_{13} \sin 2 \theta_{12} \tan \theta_{23} ) ~\delta m^2_{21}]\\
 \langle m^2_\tau \rangle_{12} &=& 
 \frac{1}{2}[m^2_2+m^2_1+(\cos 2 \theta_{12}+2 \cos \delta \sin \theta_{13} \sin 2 \theta_{12} \cot \theta_{23} ) ~\delta m^2_{21}], \nonumber
\end{eqnarray}
where the $\tau$-flavor flavor average is given for completeness only.

It is now obvious that $\nu_e$ and $\nu_\mu$ disappearance experiments
measure {\it different} $\delta m^2_{\rm eff}$'s.  In fact the three
$\delta m^2_{\rm eff}$ are\footnote{The effective atmospheric mass squared difference for the muon channel has been discussed
 in ref.~\cite{WIN03}.}
  \begin{eqnarray}
    \delta m^2_{\rm eff}|_e&=& \cos^2 \theta_{12} \delta m^2_{31}+\sin^2 \theta_{12}\delta m^2_{32}
    \label{meffe} \\
    \delta m^2_{\rm eff}|_\mu&=& \sin^2 \theta_{12} \delta m^2_{31}+\cos^2 \theta_{12}\delta m^2_{32}
    +\cos \delta \sin \theta_{13} \sin 2 \theta_{12} \tan \theta_{23}\delta m^2_{21}
\label{meffm}    \\
    \delta m^2_{\rm eff}|_\tau&=& \sin^2 \theta_{12} \delta m^2_{31}+\cos^2 \theta_{12}\delta m^2_{32}
    -\cos \delta \sin \theta_{13} \sin 2 \theta_{12} \cot \theta_{23}\delta m^2_{21}. \label{mefftau}
    \end{eqnarray}

In Fig.~\ref{prob} we show the survival probability in the $\bar{\nu}_e$ and $\nu_\mu$
disappearance channels using three different choices of the 
atmospheric $\delta m^2$
whose sign flip, with constant magnitude, changes the hierarchy from normal to inverted.  
When we use  $\delta m^2_{eff}|_\alpha$ for the $\alpha$ flavor, the change in the survival probability 
is very small when we flip the hierarchy i.e. the magnitude of this $\delta m^2_{eff}$ is insensitive
to which hierarchy nature has chosen.
Although $\delta m^2_{31}$ ($\delta
m^2_{32}$) works better for  $\bar{\nu}_e$ ($\nu_\mu$) disappearance experiments neither choice is
as good as $\delta m^2_{eff}$.
Thus, in summary,  $\delta m^2_{eff}|_e$, Eqn[\ref{meffe}], is the atmospheric $\delta m^2$ measured by $\bar{\nu}_e$ disappearance
experiments and $\delta m^2_{eff}|_\mu$, Eqn[\ref{meffm}], is the atmospheric $\delta m^2$ measured by $\nu_\mu$ disappearance
experiments upto corrections of ${\cal{O}}(\delta m^2_{21}/\delta m^2_{32})^2$.

%The flavor dependence can be seen in Fig.~\ref{prob} where we
%observe that to take $\delta m^2_{31}$ as the atmospheric $\delta m^2$
%is more suitable for reactor experiments, while the choice $\delta
%m^2_{32}$ is better for accelerator experiments. 
%Nevertheless, $\delta m^2_{\rm eff}\vert_\alpha$ for the $\alpha$-flavor is, we claim, always the best %choice.  

\begin{figure}
\centering\leavevmode
\vglue -1.0cm
%\hglue 2.0cm
\includegraphics[width=12.cm]{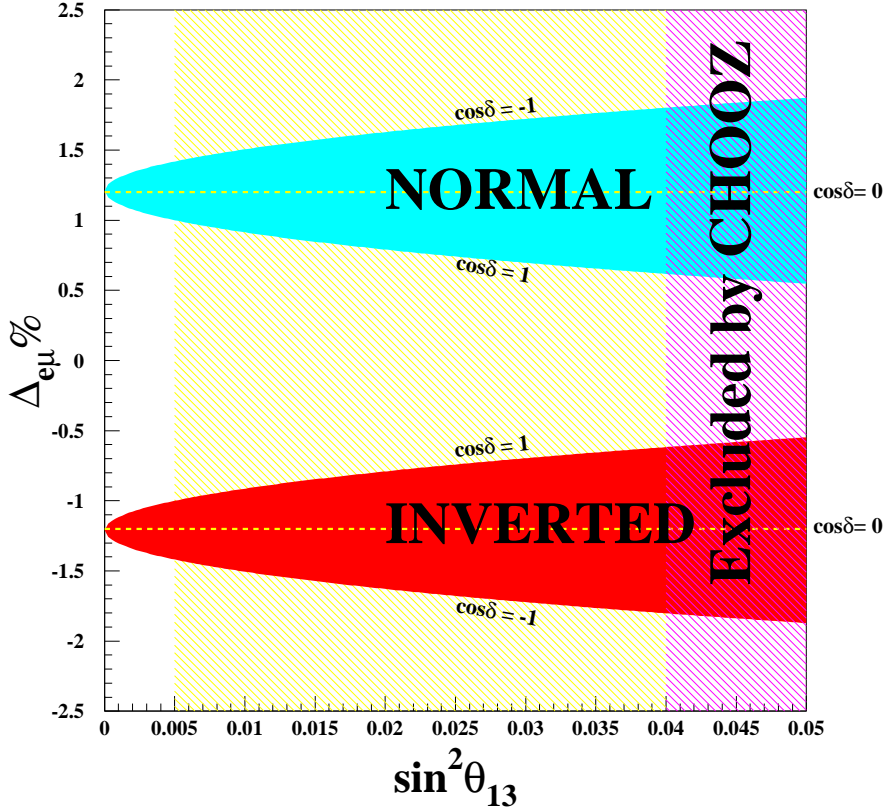}
\vglue -0.5cm
\caption{The fractional difference of the electron and muon neutrino
effective atmospheric $\delta m^2$,  $\Delta_{e\mu} \equiv  (\vert \delta m^2_{\rm eff}\vert_e -
  \vert\delta m^2_{\rm eff}\vert_\mu)/ \vert \delta m^2_{\rm eff}\vert$, as a function of
  $\sin^2\theta_{13}$ for the normal and inverted hierarchies showing the dependence on $\cos \delta$.
  The vertical scale varies linearly with the not so well known ratio of 
  $\delta m^2_{21}/\delta m^2_{32}$; here we have used
  $\delta m^2_{21}=8.0 \times 10^{-5} ~{\rm eV}^2$ and
  $\delta m^2_{32}=2.5 \times 10^{-3} ~{\rm eV}^2$.
  In a reactor $\bar{\nu}_e$ disappearance experiment, precision measurement
  of the effective atmospheric $\delta m^2_{eff} \vert_e$ is probably very 
  difficult unless $\sin^2 \theta_{13} > 0.005$.
  }
\label{ratio}
\end{figure}

  Whether the absolute value of $ \delta m^2_{\rm eff}|_e$ is larger
  or smaller than the absolute value of $ \delta m^2_{\rm eff}|_\mu$
  depends on whether $|\delta m^2_{31}|$ is larger or smaller than
  $|\delta m^2_{32}| $.  The relative magnitude of these two $\delta
  m^2$ is determined by whether the mass squared of the 3-state is
  larger or smaller than the mass squared of the 1- and 2-states, i.e.
  by the neutrino mass hierarchy.  It is easy to show that the difference
  in the absolute value of the e-flavor and $\mu$-flavor $\delta
  m^2_{\rm eff}$'s is given by 
{\mathversion{bold}
\begin{eqnarray}
\vert \delta m^2_{\rm eff}\vert_e  - \vert\delta m^2_{\rm eff}\vert_\mu 
= \pm \delta m^2_{21} (\cos 2\theta_{12}
-\cos \delta \sin \theta_{13} \sin 2 \theta_{12} \tan \theta_{23}),
\label{id1}
\end{eqnarray}
} 
where the $+$ sign ($-$ sign) is for the normal (inverted)
hierarchy.  Thus by precision measurements of both of these $\delta
m^2_{\rm eff}$ one can determine the hierarchy and possibly even $\cos
\delta$ at very high precision.  
This identity, Eqn.[\ref{id1}], is the principal observation
of this paper.

In Fig.~\ref{ratio}, we show the fractional difference in the effective 
atmospheric $\delta m^2$
for the normal and inverted hierarchy, as a
function of $\sin^2\theta_{13}$. For the normal hierarchy,
independently of $\delta$, this normalized ratio is always positive,
while for the inverted hierarchy, it is always negative.  While the
size of difference between the two hierarchies is smallest for $\cos
\delta = 1$, for this value of $\delta$, the difference between the
two hierarchies increases as $\sin^2 \theta_{13}$ goes to zero, as can
be seen from Eqn[\ref{id1}].

\begin{figure}[tbh]
\centering\leavevmode
\vglue -1.0cm
%\hglue 2.0cm
\includegraphics[width=12.cm]{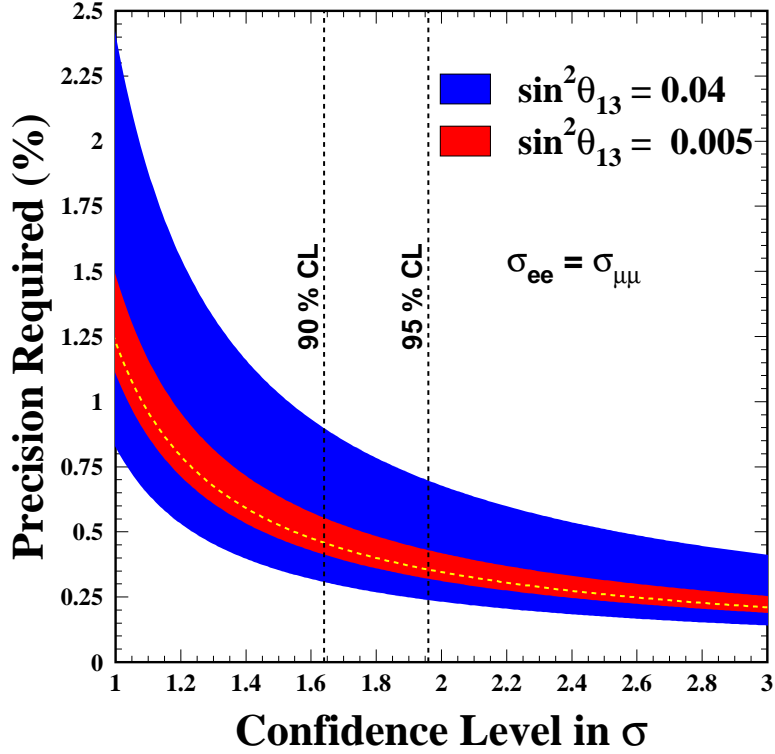}
\vglue -0.5cm
\caption{The required percentage precision need to determine the neutrino mass hierarchy
verses the confidence level of that determination.
Here we have assumed both 
effective atmospheric $\delta m^2$ are measured with the same precision,
$\sigma_{ee}=\sigma_{\mu \mu}$.
The cosine of the CP violating phase is varied from +1 (bottom) through 0 (dashed
line) to -1 (top).
Again, the vertical scale varies linearly with the not so well known ratio of 
  $\delta m^2_{21}/\delta m^2_{32}$. For this figure we have used 0.032, 
  the same as in Fig.~\ref{ratio}.
  }
\label{prec}
\end{figure}

What kind of precision is required? Given that
\begin{eqnarray}
{\delta m^2_{21}\over \delta m^2_{32}}  & \approx & \frac{1}{30} \quad \quad
{\rm and} \quad \quad \cos 2 \theta_{12} \approx  0.38,
\end{eqnarray}
the difference in the magnitude of the two effective atmospheric
$\delta m^2$ is 1 to 2\%. 
Currently, the uncertainty on the size of this difference is dominated by the experimental uncertainty 
on the ratio of the  solar to atmospheric $\delta m^2$'s.
To determine the hierarchy we need to
determine whether $\vert \delta m^2_{\rm eff}\vert_e$ is larger,
normal hierarchy, or smaller, inverted hierarchy, than $\vert\delta
m^2_{\rm eff}\vert_\mu$.  Thus determining the hierarchy with a
confidence level near 90\% one needs to measure {\it both} $\delta
m^2_{\rm eff}$ to better than one per cent precision.  These are very
challenging levels of precision for atmospheric $\delta m^2$ measurements
both within a given experiment and between two different experiments.
In Fig.\ref{prec}
we have calculated the required precision as function of the C.L., measured in sigmas,
assuming that the two experiments have the same \% precision.
From this figure we see that for a 90\% C.L.
determination of the hierarchy one would  require $\sim$0.5\%
precision on {\it both} $\delta m^2_{eff}$ measurements.
Achieving such precision will require significant innovation.  

So far our discussion has only been in vacuum. What about matter
effects? How much do they shift the first extrema?  For the $\nu_e$
disappearance channel the shift in the extrema is proportional to (aL)
where $a= \displaystyle G_F N_e/\sqrt{2} \approx (4000~{\rm
km})^{-1}$.  Thus the expected shift is less than 0.1\% for a baseline
of a few kilometers.  The size of this shift has been confirmed by a
numerical calculation.  For the $\nu_\mu$ disappearance channel again
the shift in the extrema is again proportional to (aL) but here the
baseline could go up to 1000~km.  However the coefficient in front of
(aL) is proportional to $\sin^2 2\theta_{13}$ and $\cos
2\theta_{23}/\cos^2 \theta_{23}$ both of which are small numbers.
Using an energy so that the first minimum occurs at 1000~km, we have
calculate numerically the size of the shift assuming $\sin^2
2\theta_{13}$ is at the Chooz bound and found that the maximum shift
is 0.4\%.  This maximum shift occurs when $\theta_{23}$ is as larger
as is allowed by atmospheric neutrino data.  If $\sin^2\theta_{23}$
and/or $\sin^2\theta_{13}$ are smaller than these maximum values then
the shift is smaller.  Also the shift at baselines smaller than
1000~km are proportionally smaller.  Therefore, we conclude that in
general matter effects can be safely ignored, or corrected for, in $\nu_\mu$ disappearance
experiments whose baseline is less than 1000~km.

In summary we have demonstrated that high precision measurements of
the effective atmospheric $\delta m^2$ in both the $\bar{\nu}_e \to
\bar{\nu}_e$ (reactor) and $\nu_\mu \to \nu_\mu$ (long baseline
accelerator) channels can determine the neutrino mass hierarchy
independent of matter effects. The sign of the difference determines
the hierarchy.  For any reasonable confidence level determination the
precision required in {\it both} channels is a very challenging
fraction of 1\%.  The next generation of long baseline experiments
such as T2K~\cite{jhf} and NO$\nu$A~\cite{nova} estimate their
precision on the effective atmospheric $\delta m^2$ at 2\%.  However,
so far there has been no physics reason to push this to a precision
measurement.  For the reactor channel the emphasis so far has been on
the observation of non-zero $\theta_{13}$~\cite{whitepaper}, very
little effort has been made on a precision determination of the
effective atmospheric $\delta m^2$. This kind of precision, can perhaps     
be achieved in beta beam facility~\cite{beta}.  
We realize that to make these
measurements to the precision suggested is very challenging
experimentally. 
However we encourage our experimental colleagues to
give this some thought especially since this method has a different
dependence on the unknown CP violating phase, $\cos \delta$ versus $\sin \delta$, 
compared with long baseline experiments.

While we were completing this manuscript, ref.\cite{boris} appeared which
discusses the physics of this possibility in a pure 3-flavor frame work as well
as discussing other possible ways of determining the hierarchy.
 
 \vspace{-0.3cm}
 \begin{acknowledgments} 
 \vspace{-0.3cm}
  This work was supported by Funda\c{c}\~ao de Amparo \`a Pesquisa do
  Estado de S\~ao Paulo (FAPESP), Conselho Nacional de Ci\^encia e
  Tecnologia (CNPq). 
  %and by the National Science Foundation under Grant No.  PHYYY-XXX. 
  Fermilab is operated under DOE contract DE-AC02-76CH03000.
 Two of us (H.N. and R.Z.F.) are grateful for the
  hospitality of the Theory Group of the Fermi National Accelerator
  Laboratory during the summer of 2004, where most of this work was completed.
   \end{acknowledgments}

\vspace{1cm}

%%%%%%%%%%%%%%%%%%%%%%%%%%%


\begin{thebibliography}{99}
%%%%%%%%%%%%%%%%%%%%%%%%%%%
%%%%%%%%%% Experiments 
\bibitem{SNO} B. Aharmim {\it et al.} [SNO Collaboration],  
arXiv:hep-ex/0502021.
\bibitem{KamLAND} T. ~Araki {\it et al.} [KamLAND Collaboration],
  arXiv:hep-ex/0406035.
\bibitem{sinsq}
 O.~Mena and S.~J.~Parke,
  %``Unified graphical summary of neutrino mixing parameters,''
  Phys.\ Rev.\ D {\bf 69}, 117301 (2004)
  [arXiv:hep-ph/0312131].
  %%CITATION = HEP-PH 0312131;%%
\bibitem{SK_solar} M. B. Smy {\it et al.} [Super-Kamiokande Collaboration],
Phys. Rev. D {\bf 69}, 011104 (2004). 
\bibitem{solar_other} 
B.~T.~Cleveland {\it et al.},
%``Measurement Of The Solar Electron Neutrino Flux With The Homestake  Chlorine Detector,''
Astrophys.\ J.\  {\bf 496}, 505 (1998);
%%CITATION = ASJOA,496,505;%%
%
J.~N.~Abdurashitov {\it et al.}  [SAGE Collaboration],
%``Measurement of the solar neutrino capture rate with gallium metal,''
Phys.\ Rev.\ C {\bf 60}, 055801 (1999)
[arXiv:astro-ph/9907113];
%%CITATION = ASTRO-PH 9907113;%%
%
W.~Hampel {\it et al.}  [GALLEX Collaboration],
%``GALLEX solar neutrino observations: Results for GALLEX IV,''
Phys.\ Lett.\ B {\bf 447}, 127 (1999);
%%CITATION = PHLTA,B447,127;%%
%C. Cattadori, {\it Results from Radiochemical
 % Solar Neutrino Experiments}, XXIth International Conference on
 % Neutrino Physics and Astrophysics (Neutrino 2004), Paris, June
 % 14-19, 2004.
\bibitem{SK_atm} Y. Fukuda {\it et al.} [Super-Kamiokande
  Collaboration], Phys. Rev. Lett. {\bf 81}, 1562 (1998); Y. Ashie
  {\it et al.}, Phys. Rev. Lett. {\bf 93}, 101801 (2004);
  arXiv:hep-ex/0501064.
\bibitem{K2K} M. H. Ahn {\it et al.} [K2K Collaboration], Phys. Rev.
  Lett. 90, 041801 (2003).
%%%%%%%%%% Mass Hierarchy w/Matter effects in conventional LBL 
\bibitem{Chooz}
  M.~Apollonio {\it et al.}  [CHOOZ Collaboration],
  %``Initial results from the CHOOZ long baseline reactor neutrino  oscillation
  %experiment,''
  Phys.\ Lett.\ B {\bf 420}, 397 (1998)
  [arXiv:hep-ex/9711002];
%%CITATION = HEP-EX 9711002;%%
%\cite{Apollonio:1999ae}
%\bibitem{Apollonio:1999ae}
%  M.~Apollonio {\it et al.}  [CHOOZ Collaboration],
  %``Limits on neutrino oscillations from the CHOOZ experiment,''
  Phys.\ Lett.\ B {\bf 466}, 415 (1999)
  [arXiv:hep-ex/9907037].
%%CITATION = HEP-EX 9907037;%%
\bibitem{lbl} H.~Minakata and H.~Nunokawa, 
%``Exploring neutrino mixing with low energy superbeams,''
JHEP {\bf 0110}, 001 (2001)
[arXiv:hep-ph/0108085];
%%CITATION = HEP-PH 0108085;%%
V.~Barger, D.~Marfatia and K.~Whisnant,
%``Breaking eight-fold degeneracies in neutrino CP violation, mixing, and
%mass hierarchy,''
Phys.\ Rev.\ D {\bf 65}, 073023 (2002) 
[arXiv:hep-ph/0112119];
V.~Barger, D.~Marfatia and K.~Whisnant,
%``Off-axis beams and detector clusters: Resolving neutrino parameter
%degeneracies,''
Phys.\ Rev.\ D {\bf 66}, 053007 (2002)
[arXiv:hep-ph/0206038];
%%CITATION = HEP-PH 0206038;%%
P.~Huber, M.~Lindner and W.~Winter,
%``Synergies between the first-generation JHF-SK and NuMI superbeam
%experiments,''
Nucl.\ Phys.\ B {\bf 654}, 3 (2003)
[arXiv:hep-ph/0211300];
%%CITATION = HEP-PH 0211300;%%
O.~Mena and S.~J.~Parke,
%``Untangling CP violation and the mass hierarchy in long baseline
%experiments,''
Phys.\ Rev.\ D {\bf 70}, 093011 (2004)
[arXiv:hep-ph/0408070].
%%CITATION = HEP-PH 0408070;%%

%%%%%%%%%% Mass Hierarchy w/Matter effects in SN
\bibitem{sn}V. Barger, P. Huber and D. Marfatia, 
arXiv:hep-ph/0501184; C. Lunardini and A. Yu. Smirnov, 
JCAP {\bf 0306}, 009 (2003); A. S. Dighe {\it et al.}, 
JCAP {\bf 0306}, 059 (2003); A.~S.~Dighe and A.~Y.~Smirnov,
%``Identifying the neutrino mass spectrum from the neutrino burst from a
%supernova,''Phys.\ Rev.\ D {\bf 62}, 033007 (2000)
[arXiv:hep-ph/9907423].
%%CITATION = HEP-PH 9907423;%%
H.~Minakata and H.~Nunokawa,
%``Inverted hierarchy of neutrino masses disfavored by supernova 1987A,''
Phys.\ Lett.\ B {\bf 504}, 301 (2001)
[arXiv:hep-ph/0010240];
C.~Lunardini and A.~Y.~Smirnov,
%``Supernova neutrinos: Earth matter effects and neutrino mass spectrum,''
Nucl.\ Phys.\ B {\bf 616}, 307 (2001)
[arXiv:hep-ph/0106149].

%%%%%%%%%% Mass Hierarchy w/Matter effects in muon collider
\bibitem{muon} T. Adams {\it et al.}, ``E1 working group summary:
  Neutrino factories and muon collider'', in {\it Proc. of the
    APS/DPF/DPB Summer Study on the Future of Particle Physics
    (Snowmass 2001)''}, ed. N. Graf, eConf {\bf C010630}, E1001 (2001)
    [arXiv:hep-ph/0111030]; C.~Albright {\it et al.},
%``Physics at a neutrino factory,''
arXiv:hep-ex/0008064.
%%CITATION = HEP-EX 0008064;%%
A.~Cervera, A.~Donini, M.~B.~Gavela, J.~J.~Gomez Cadenas,
P.~Hernandez, O.~Mena and S.~Rigolin,
%``Golden measurements at a neutrino factory,''
Nucl.\ Phys.\ B {\bf 579}, 17 (2000) [Erratum-ibid.\ B {\bf 593}, 731
(2001)] [arXiv:hep-ph/0002108]; M.~M.~Alsharoa {\it et al.}  [Muon
Collider/Neutrino Factory Collaboration],
%``Recent progress in neutrino factory and muon collider research within the
%Muon collaboration,''
Phys.\ Rev.\ ST Accel.\ Beams {\bf 6}, 081001 (2003)
[arXiv:hep-ex/0207031];
M.~Apollonio {\it et al.},
%``Oscillation physics with a neutrino factory. ((G)) ((U)),''
arXiv:hep-ph/0210192.
%%%%%%%%%%%%%%%%%%%%%%%%%
\bibitem{Petcov}
   S.~T.~Petcov and M.~Piai,
  %``The LMA MSW solution of the solar neutrino problem, inverted neutrino  mass
  %hierarchy and reactor neutrino experiments,''
  Phys.\ Lett.\ B {\bf 533}, 94 (2002)
  [arXiv:hep-ph/0112074];
  %%CITATION = HEP-PH 0112074;%%
S.~Choubey, S.~T.~Petcov and M.~Piai,
  %``Precision neutrino oscillation physics with an intermediate baseline
  %reactor neutrino experiment,''
  Phys.\ Rev.\ D {\bf 68}, 113006 (2003)
  [arXiv:hep-ph/0306017].
  %%CITATION = HEP-PH 0306017;%%
  %%%%%%%%%%%%%%%%%%%%%%%%%%%%%%%%%
\bibitem{MNS}
 Z.~Maki, M.~Nakagawa and S.~Sakata,
  %``Remarks On The Unified Model Of Elementary Particles,''
  Prog.\ Theor.\ Phys.\  {\bf 28}, 870 (1962);
  %%CITATION = PTPKA,28,870;%%
We use the standard representation this matrix, see for example ref.\cite{sinsq}.
\bibitem{WIN03}
S. Parke WIN03
http://conferences.fnal.gov/win03/;
  N.~Okamura,
  %``Effect of the smaller mass-squared difference for the long base-line
  %neutrino experiments,''
  arXiv:hep-ph/0411388.
  %%CITATION = HEP-PH 0411388;%%
%%%%%%%%%% JHF Proposal
\bibitem{jhf} Y. Itow {\it et al.}, arXiv:hep-ex/0106019.
For an updated version, see:
http://neutrino.kek.jp/jhfnu/loi/loi.v2.030528.pdf
%%%%%%%%%% NOVA Proposal
\bibitem{nova} I. Ambats {\it et al.}, NO$\nu$A: Proposal to build an
  off-axis detector to study $\nu_\mu \to \nu_e$ oscillations in the
  NUMI beamline. FERMILAB-PROPOSAL-0929.
%%%%%%%%%% Reactor theta_13 - White paper
\bibitem{whitepaper} K.~Anderson {\it et al.}, White Paper Report on
  Using Nuclear Reactors to Search for a Value of $\theta_{13}$,
  arXiv:hep-ex/0402041.
\bibitem{beta}
  P.~Zucchelli,
%``A novel concept for a anti-nu/e / nu/e neutrino factory: The beta beam,''
  Phys.\ Lett.\ B {\bf 532} (2002) 166; J.~Burguet-Castell, D.~Casper,
  J.~J.~Gomez-Cadenas, P.~Hernandez and F.~Sanchez,
%``Neutrino oscillation physics with a higher gamma beta-beam,''
  Nucl.\ Phys.\ B {\bf 695}, 217 (2004)  [arXiv:hep-ph/0312068];
  A.~Donini, E.~Fernandez-Martinez, P.~Migliozzi, S.~Rigolin and L.~Scotto Lavina,
%``Study of the eightfold degeneracy with a standard beta-beam and a
%super-beam facility,''
Nucl.\ Phys.\ B {\bf 710}, 402 (2005)
[arXiv:hep-ph/0406132].

\bibitem{boris} A. de Gouvea, J. Jenkins and B. Kayser, arXiv:hep-ph/0503079.

\end{thebibliography}
\end{document}